\documentclass[final]{IEEEtran}

\usepackage{graphicx}
\usepackage{amsmath}
\usepackage{amssymb}
\usepackage{cite}

\newcommand{\join}{\widehat{\hspace{9pt}}}
\newcommand{\cherries}{\mathit{Ch}}
\newcommand{\R}{\mathbb{R}}

\newcommand{\act}{\phi}
\newcommand{\op}{\star}
\newcommand{\ceq}{\stackrel{c}{=}}

\newcommand{\Symm}{\mathrm{Symm}}
\newcommand{\betast}{\ensuremath{\psi}}

\newtheorem{prop}{Proposition}

\newtheorem{defn}{Definition}

\title {Optimization over a class of tree shape statistics}

\author{Frederick A. \ Matsen \\
Biomathematics Research Centre \\
University of Canterbury \\
Private Bag 4800 \\
Christchurch \\
New Zealand \\
phone: +64 3 364 2987 x7431\\
fax: +64 3 364 2587\\
e.matsen@math.canterbury.ac.nz \\
http://www.math.canterbury.ac.nz/matsen/}

\begin{document}

\maketitle

\begin{abstract}
  Tree shape statistics quantify some aspect of the shape of a
  phylogenetic tree. They are commonly used to compare reconstructed
  trees to evolutionary models and to find evidence of tree
  reconstruction bias. Historically, to find a useful tree shape
  statistic, formulas have been invented by hand and then evaluated
  for utility. This article presents the first method which is capable
  of optimizing over a class of tree shape statistics, called
  \emph{Binary Recursive Tree Shape Statistics (BRTSS)}. After
  defining the BRTSS class, a set of algebraic expressions is defined
  which can be used in the recursions. The tree shape statistics
  definable using these expressions in the BRTSS is very general, and
  includes many of the statistics with which phylogenetic researchers
  are already familiar. We then present a practical genetic algorithm
  which is capable of performing optimization over BRTSS given any
  objective function. The chapter concludes with a successful
  application of the methods to find a new statistic which indicates a
  significant difference between two distributions on trees which were
  previously postulated to have similar properties.  
\end{abstract}


Tree shape statistics are numerical summaries of some aspect of the
shape of a phylogenetic tree. The first tree shape statistic was the
$\bar{N}$ of Sackin \cite{Sackin1972:225}, where the explicit goal was
to numerically describe the \emph{balance} of a tree, which is the
degree to which daughter subtrees of internal nodes are of similar or
different size. Trees which are balanced have smaller $\bar{N}$ than
do trees which are imbalanced. Many other tree shape statistics
followed, all quantifying balance; a review of this literature can be
found in the excellent review article by Mooers and Heard
\cite{Mooers1997:31}.

The next important step in tree shape theory was made by Kirkpatrick and
Slatkin \cite{Kirkpatrick1993:1171} who wondered which statistics were
the most powerful to distinguish between the so-called ERM and PDA
distributions on trees. The statistics which they chose to rate
included most of the statistics available in the literature at that
time. Their article is among the most influential in the area of tree
shape, with over 60 citations as of March 2006 (ISI Web of Knowledge
search, http://portal.isiknowledge.com/).

The article by Kirkpatrick and Slatkin marked a philosophical
shift from the idea of a tree shape statistic as a purely descriptive
device to that of a mapping which can be used in a statistical
fashion. Their work was continued more recently by Agapow and Purvis
\cite{Agapow2002:866} who took seven tree shape statistics from the
literature and one of their own, then tested them for power in
distinguishing several different models. They then made general
recommendations for which statistics to use.

The next step for the Kirkpatrick-Slatkin methodology needs
to overcome two limitations. First, the statistics which are tested
are typically invented ``by hand'' and so are limited by the ingenuity
of individual authors. Second, general recommendations may not be
sufficient for all situations in which tree shape statistics are
useful. For example, although a statistic such as Colless' index
\cite{Colless1982:100} has lots of power in the Kirkpatrick-Slatkin
and Agapow-Purvis scenarios, it has low power to distinguish between
two distributions which have similar overall balance \cite{matsen05a}.

This paper presents a methodology which enables, for the first time,
direct optimization over tree shape statistics. First, we present a
recursive framework and a class of algebraic expressions which can be
used to define tree shape statistics in a natural way. These
statistics are a large and varied family which include most of the
present-day tree shape statistics. Second, this paper presents a
practical genetic algorithm which, given an objective function, can be
applied to produce high-performance tree shape statistics.

For the purpose of this paper, \emph{tree} refers to a finite rooted
bifurcating tree without leaf labels or edge lengths.

\section{Binary recursive tree shape statistics}
\subsection{Definition and examples}

This section defines binary recursive tree shape statistics (BRTSS)
which form the framework over which the optimization algorithms
operate. The starting observation for the definition is that many extant
statistics are constructed with reference to their values on subtrees.
For example, the number of leaves of a tree can be calculated 
recursively by summing the number of leaves of its two subtrees.
Using $X \join Y$ to signify a tree with $X$ and $Y$ as subtrees, one
can write this statement as
\[
l(X \join Y) = l(X) + l(Y).
\]

One can write a tree shape statistic of this sort by specifying a
``recursion'' $\rho$ and a ``base case'' $\lambda$,
\begin{equation}
\label{eq:simple_brts_eval}
s(T) = \left \{ 
  \begin{array}{ll}
    \rho ( s(X), s(Y) ) & \mathrm{if}\ T = X \join Y \\
    \lambda & \mathrm{if}\ T\ \hbox{is a leaf}
  \end{array}
  \right.
\end{equation}
Because $X \join Y = Y \join X$, the resulting $s$ is well defined
only if $\rho$ is symmetric, i.e. if $\rho(x,y) = \rho(y,x)$. In the
above
notation the number of leaves of a tree can be written as a recursive
tree shape statistic with $\lambda = 1 $ and $\rho(x,y) = x+y$.
Another example of this sort of statistic is the maximal depth of the
tree, for which $\lambda = 0$ and $\rho(x,y) = 1+\max(x,y)$.

A remarkable number of useful statistics can be achieved by varying
$\rho$ and $\lambda$. However, considerably more can be written using
several mutually recursive statistics. For example, perhaps the most
popular tree shape statistic is the ``Colless index'' $I_c$
\cite{Mooers1997:31} \cite{Colless1982:100}. This index (without a
normalizing factor) sums for each internal node the absolute value of
the difference between the number of leaves of the two daughter
subtrees of that
internal node. It can be written as follows:
\[
I_c(T) = \left \{ 
  \begin{array}{ll}
     I_c(X) + I_c(Y) + |l(X) - l(Y)| & \mathrm{if}\ T = X \join Y \\
    0 & \mathrm{if}\ T\ \hbox{is a leaf.}
  \end{array}
  \right.
\]
This version of $I_c$ is constructed from two real numbers (the base
cases) and two functions symmetric in $X$ and $Y$ (the recursions).
This leads to the definition of a \emph{BRTSS}.

\begin{defn}
  \label{def:BRTSS}
  A BRTSS of length $n$ is an ordered pair
  $(\mathbf{\lambda},\mathbf{\rho})$ where
    $\mathbf{\lambda} \in \R^n$ and
    $\mathbf{\rho}$ is an $n$-vector of $ \Symm^2(\R^n)
      \rightarrow \R$ maps.
\end{defn}
In the definition, $\Symm^2(\R^n)$ denotes the symmetric product of $\R^n$
with itself. The condition
that the $\rho_i$ map from the symmetric product is equivalent to
saying that they map $\R^{2n} \rightarrow \R$ and are invariant under
the action exchanging the $x_i$'s and the $y_i$'s, i.e. for any $j$
\begin{equation}
  \label{eq:symmetry}
      \rho_j(x_1, \ldots, x_n, y_1, \ldots, y_n) = 
      \rho_j(y_1, \ldots, y_n, x_1, \ldots, x_n).
\end{equation}

A BRTSS is evaluated on a tree by a generalization of
(\ref{eq:simple_brts_eval}). Recursively define the $s_i$ by 
\begin{equation}
\label{eq:brts_eval}
s_i(T) = \left \{ 
  \begin{array}{l}
    \rho_i ( s_1(X), \ldots, s_n(X), s_1(Y), \ldots, s_n(Y) ) \\
    \lambda_i 
  \end{array}
  \right.
\end{equation}
where the first case is used in case
 $T = X \join Y$, and the second if $T$ is a leaf.
The final value of the BRTSS on $T$ is simply defined to be $s_1(T)$.
The symmetry property of the $\rho_i$ imply that (\ref{eq:brts_eval})
is well defined. For this paper, $x_i$ (resp. $y_i$) will be used for
the value of $s_i$ on the subtree $X$ (resp. $Y$).

The Colless index (without the normalizing factor) can now be written
as a BRTSS of length 2 with the base cases $\lambda_1 = 0$,
$\lambda_2 = 1$, and the two recursions
\begin{eqnarray*}
  \rho_1(x_1,x_2,y_1,y_2) & = & x_1 + y_1 + |x_2 - y_2| \\
  \rho_2(x_1,x_2,y_1,y_2) & = & x_2 + y_2.
\end{eqnarray*}
The second recursion $\rho_2$ for $I_c$ simply totals the value of
$s_2$ applied to subtrees $X$ and $Y$. With the base case
$\lambda_2 = 1$, this implies that $s_2$ gives the number of leaves of
the tree as before. The first recursion $\rho_1$ adds the absolute
value of the difference of $s_2$ applied to the subtrees to the sum of
the values of $s_1$ applied to the subtrees. This is indeed the
(un-normalized) Colless index $I_c$ as described above.

The BRTSS formulation can be used to define many tree shape statistics
from the literature with simple recursive functions $\rho$.  For
example, we show here how to define the number of two leaf subtrees of
a tree (called the number of ``cherries'' \cite{mckenzie-steel}), and
un-normalized versions of Sackin's $\bar{N}$ \cite{Sackin1972:225},
and Shao and Skokal's $B_1$ \cite{Shao1990:266}. The latter two can be
defined as follows. Let $\mathcal{I}$ denote the internal nodes and
$r$ denote the root of a tree.  For $i \in \mathcal{I}$, let $N_i$ be
the number of leaves of the subtree subtended by $i$. For a node $j
\in \mathcal{I} - \{r\}$, let $M_j$ be the maximal depth of the
subtree with $j$ as the root. Then \begin{equation}
  \label{eq:classical_defs} \bar{N} = \sum_{i \in \mathcal{I}} N_i
  \hspace{1cm} B_1 = \sum_{j \in \mathcal{I} - \{r\}} M_j^{-1}.
\end{equation} The above formulas for the number of cherries
$\cherries$, $\bar{N}$, and $B_1$, respectively, can be written in
BRTSS form as\footnote{For BRTSS evaluation we will use the convention
that 0/0 = 0. This allows for more flexibility in the use of
division.}

\begin{IEEEeqnarray*}{lll}
\cherries \, & : & ((0,1), (x_1+y_1+I(x_2+y_2,2),\ x_2 + y_2))\\
\bar{N} & : & ((0,1), (x_1+y_1+x_2+y_2,\ x_2 + y_2))\\
B_1 & : & \left((0,0), \left(x_1 + y_1 + \frac{1-I(x_2,0)}{x_2} \right. \right . \\
& & \hspace{2.7cm} + \left. \left. \frac{1-I(y_2,0)}{y_2}, 1+\max(x_1,y_1) \right)\right) \\
\end{IEEEeqnarray*}

In the above $I$ denotes the binary indicator function, i.e. $I(a,b)$
is one if $a=b$ and zero otherwise.  We note that formulae for
$I_c$ and $\bar{N}$ similar to the above have been published
previously in \cite{blum-theory}.

The emphasis in this paper will be on BRTSS with reasonably simple
$\rho$, however, it is true that any mapping of trees to the real line
can be written as a BRTSS using sufficiently complex $\rho$.  Begin by
enumerating the (countable) set of trees and define $s_2(T)$ to
be the number of a given tree $T$.  This can be written recursively by
setting $\rho_2 (x_1, y_1, x_2, y_2)$ to be the number of the
tree which has the trees numbered $x_2$ and $y_2$ as subtrees.  The
function $\rho_1$ simply gives the desired value of the statistic on
the tree composed of the two subtrees numbered $x_2$ and $y_2$.  This
statistic is a BRTSS by definition.

\subsection{Verifiably symmetric algebraic expressions}
\label{sec:verif-symmetric}

The primary aim of this paper is to demonstrate a system capable of
optimizing over a class of tree shape statistics.  The previous
section defined the BRTSS class, which defines a tree shape statistic
in a natural way given a real vector $\mathbf{\lambda} \in \R^n$ and
$\mathbf{\rho}$, an $n$-vector of $\Symm^2(\R^n) \rightarrow \R$ maps.
The promised optimization will proceed by modifying the
$\mathbf{\lambda}$ and $\mathbf{\rho}$ vectors.  Optimizing over
$n$-dimensional real vectors is a classical subject, however
optimization over such symmetric maps generally is not.  Any class of
such $\Symm^2(\R^n) \rightarrow \R$ maps could in principle be used as
a set for enumeration and optimization, however a balance must be
struck between ease of optimization and generality.  For instance, one
could easily use symmetric linear functions as the underlying
recursions and adjust the coefficients in order to find statistics
with desirable properties.  However, this rather restrictive class
would exclude all of the above BRTSS except for $l$ and $\bar{N}$. 

The purpose of this section is to introduce a subset of the
$\Symm^2(\R^n) \rightarrow \R$ functions which is quite general though
sufficiently simple to be the underlying population for a genetic
algorithm.  This subset is functions induced by a class of
algebraic expressions with certain allowed operations and
operands.  The challenge lies in ensuring the
symmetry property (\ref{eq:symmetry}). 

The basic idea of this class of algebraic expressions, which will be
called \emph{verifiably symmetric algebraic expressions}, is simple: we
constrain the algebraic expressions to remain the same (up to the
order of operands of commutative operations) after exchanging the
$x_i$ for the $y_i$.  For example,
exchanging $x_1$ for $y_1$ in $x_1 + y_1$ gives $y_1 + x_1$, which is
equivalent to $x_1 + y_1$ after applying the commutative rule for
addition.  Therefore, $x_1 + y_1$ is considered to be verifiably symmetric.
Similarly, $\min(x_1/y_2,y_1/x_2)$ is also verifiably symmetric
because $\min$ is a commutative binary operation. On
the other hand, the algebraic expression $0 * x_1$ is not verifiably
symmetric even though it induces a symmetric function of $x_1$ and
$y_1$.  The verifiably symmetric criterion clearly implies that the
induced functions carry the symmetry property (\ref{eq:symmetry}).

The set of finite verifiably symmetric algebraic expressions is a
convenient set over which optimization is possible. One could use
a larger set of expressions with a more
complex notion of symmetry, however, this might require consideration
of the full problem of simplification of algebraic expressions. The
simplification of algebraic expressions is a subtle field in itself
\cite{MR0239976} \cite{MR0309355} and thus generalizing the definitions might
not aid our purpose of finding a simple and useful framework for tree
shape statistics.

We now present a more rigorous version of the above definition.
\begin{defn}
An \emph{algebraic expression} on sets $C$ of constants, $V$ of variables, $U$ of unary
operations, and $B$ of binary operations is one of the following:
\begin{itemize}
  \item a constant from $C$
  \item the instantiation of a variable from $V$
  \item a unary operation from $U$ applied to an algebraic expr.
  \item a binary operation from $B$ applied to an ordered pair of algebraic expressions.
\end{itemize}
\end{defn}
A variable is different than its instantiation, as one variable may
have many distinct instantiations.  Equality of algebraic expressions
is defined recursively in the natural way.

Note that the standard rules of simplification and equivalence are not
automatic. All binary operations have an order (thus $x+y$ is not
equal to $y+x$), there is no notion of associativity, and no
simplification is done at this stage.

To construct algebraic expressions for the $\rho$ of the BRTSS, this
paper uses the integers as the set of constants and $x_1, \ldots, x_n,
y_1, \ldots, y_n$ as the set of variables, where $n$ is the length of
the BRTSS.  The standard binary operations $+,-,*$ and $/$ will be
used, as well as the binary indicator function $I$, exponentiation,
and $\max$. Of these, $+,*,I$, and $\max$ are considered to be
commutative. The unary operations used are inverse, negation, absolute
value, $\exp$, (natural) $\log$, and the symmetrization of any commutative
binary operation, which is described below.  In the following the
term ``algebraic expression'' will be used without qualification
as $C$, $V$, $U$, and $B$ have now been fixed.  Although the
definitions below do not depend on these choices, the examples will.

\begin{defn}
  Two algebraic expressions $R$ and $S$ are \emph{commutatively
    equivalent}, denoted $R \ceq S$, if $R$ can be obtained from $S$
  by changing the order of operands in commutative binary operations.
\end{defn}

Denote by $\act$ the map exchanging $x_i$ for $y_i$ in the
expressions.  Recall that this is the map used to define the symmetry
property (\ref{eq:symmetry}) of the $\rho$ in the definition of the BRTSS.

\begin{defn}
  An algebraic expression $E$ is \emph{verifiably
  symmetric} if $E \ceq \act(E)$.
\end{defn}
Given the choice of operations, examples of verifiably symmetric
algebraic expressions can be found in the above definitions of 
$\cherries$, $\bar{N}$, and $B_1$.
However, the expression
$|x_2-y_2|$ in $I_c$ is not verifiably
symmetric using our choice of operations even though the usual
algebraic simplification leads to equivalence between $|x_2-y_2|$
and its image under $\act$.  Of course, if we had decided to include
the absolute value of a difference as a
commutative binary operation in the set $B$ then $|x_2 - y_2|$ would
be considered verifiably symmetric.  Nevertheless, $|x_2-y_2|$ can be
written $\max(x_2 - y_2, y_2 - x_2)$ which is verifiably symmetric.
Therefore, $I_c$ can indeed be written as a BRTSS with verifiably
symmetric recursions.

Because of the strict hierarchy of containment set up by the
definition of algebraic expressions, the notions of sub-expression and
``smallest expression'' containing a subexpression are well defined.
\begin{defn}
  The \emph{minimal fixed expression} $M(z; E)$ of the instantiation
  $z$ of a variable appearing in a verifiably symmetric algebraic
  expression $E$ is the smallest sub-expression of $E$ containing $z$
  which is verifiably symmetric.
\end{defn}
For example, $M(x_1; 2*\log(\min(x_1,y_1))$ is $\min(x_1,y_1)$. 

A minimal fixed expression clearly cannot be a constant. Because it
is verifiably symmetric, it cannot be a variable instantiation. By
minimality, it cannot be a unary operation applied to a subexpression.
Therefore it must be of the form $E_z \op F$, where the variable 
instantiation $z$ is contained in $E_z$ and $\op$ is a binary operation.

Furthermore, since $\act(E_z \op F) \ceq E_z \op F$, either $\act(E_z)
\ceq E_z$ 
or $\act(E_z) \ceq F$. The first option is not possible:
otherwise $E_z \op F$ would not be minimal. Therefore the minimal
fixed expression $M(z; E)$ of any instantiation $z$ is of the form
$E_z \op F$ where $z$ is contained in $E_z$ and $\act(E_z) \ceq F$.
This implies further that $\op$ is commutative. Because of
the symmetry, it is possible to only store one ``side'' of the minimal
fixed expression, the other side being available through $\act$.
In the following terminology, any minimal fixed expression is
commutatively equivalent to an expression written with a
\emph{symmetrization}: 

\begin{defn}
  The \emph{symmetrization} $S_{\op}(E)$ of an expression $E$ with
  respect to a commutative binary operation $\op$\ is $E \op \act(E)$.
\end{defn}

For example, $x_1+y_1$ can be written $S_+(x_1)$, and
$\max(x_1-y_1,y_1-x_1)$ can be written $S_{\max}(x_1 - y_1)$. The
symmetrization of a binary operation is a unary operation. If every
variable instantiation in an expression is contained within at least
one symmetrization, then we will say that the expression is
\emph{completely symmetrized}. For example, $S_{*}(x_2)$ is
completely symmetrized, while $\max(x_1, S_{*} (x_2))$ is not.

Every variable instantiation in a verifiably symmetric algebraic
expression is included in a minimal fixed expression by definition,
and each such minimal expression can be written with a symmetrization
up to commutative equivalence. Therefore \begin{prop} Any verifiably
  symmetric expression is commutatively equivalent to a completely
  symmetrized algebraic expression.  \end{prop}

This simple proposition allows for a compact ``grammar'' of verifiably
symmetric algebraic expressions and a trivial way for optimization
algorithms to modify algebraic expressions while staying within the
verifiably symmetric class. The rest of this paper will consider
completely symmetrized algebraic expressions as the expressions
defining the $\rho_i$.

The value of BRTSS can be computed by free
software from http://math.canterbury.ac.nz/matsen/simmons/.

\section{Enumeration and Optimization}

\subsection{Enumeration}
\label{sec:enumeration}

With this framework it is possible to enumerate many algebraic
expressions and test them for desirable properties. This idea was
implemented as follows. First define the ``size'' of an algebraic
expression to mean the total number of operations and operands of the
expression: for example, the expression $2+x_i$ has size 3. The
symmetrization of a commutative binary operation is unary and thus adds
only one to the size. Second, select a set of constants, variables,
unary operations and binary operations for enumeration.  These can be
subsets of the complete set allowed for BRTSS recursions.

For each $k$ up to a maximal size, two lists are constructed: one of
completely symmetrized algebraic expressions and another of
non-symmetrized expressions. To construct the completely symmetrized
expressions of size $k$, all unary operations are applied to the
completely symmetrized expressions of size $k-1$, then all
symmetrizations are applied to all non-symmetrized expressions of size
$k-1$, then all binary operations are applied to all
pairs of completely symmetrized expressions of total size
$k-1$. To construct the non-symmetrized expressions of size $k$, all
unary operations are applied to the non-symmetrized expressions of
size $k-1$, then all binary operations are applied to all 
pairs of completely symmetrized and non-symmetrized expressions of
total size $k-1$, then all binary operations are applied to all
pairs of non-symmetrized expressions of total size $k-1$.

For $k=1$, the completely symmetrized algebraic expressions are taken
to be the chosen set of constants, and the non-symmetrized
algebraic expressions are instantiations of the variables. In the
present application some limited forms of simplification were
implemented to eliminate double negation and similar obvious
redundancies.

The number of statistics constructible using direct enumeration is
large. We enumerated all statistics of length one, size less than or
equal to seven, constants taken from the set $\{0,1,2\}$, variables
$x$ and $y$, and operations as in Section~\ref{sec:verif-symmetric}
except for subtraction and division, which can be expressed using
combinations of operations.
After removing those statistics which are constant on all trees on
eight leaves, 516,699 statistics remained. The number of analogous
BRTSS with length larger than one is considerably larger.

\subsection{Genetic Algorithm}

Genetic algorithms typically optimize over a very large discrete space
by maintaining a population of elements of that space and allowing
reproduction based on the value of the function to be optimized
\cite{koza}. Some notion of mutation and crossover are defined such
that the population changes over time.
Here the underlying space is taken to be the set of BRTSS with
integral $\lambda_i$ and algebraic expressions as in
\ref{sec:verif-symmetric} for the $\rho_i$. We will assume for this
section that the BRTSS under consideration have length $n$.


Standard Wright-Fisher sampling \cite{MR2026891} was applied for
reproduction. When the objective was to maximize a positive
number, the raw fitness function was simply that number. When the
objective was to minimize a number between zero and one, such as a
$p$-value, the negative of the logarithm of the objective function was
used as the raw fitness.


Two types of mutation were defined: mutation of $\lambda$ and mutation
of $\rho$. A mutation of $\lambda$ simply chooses an $i \in
\{1,\ldots,n\}$ uniformly and then adds or subtracts one from
$\lambda_i$. A mutation of $\rho$ also chooses a $\rho_i$
uniformly to mutate. A mutation of a $\rho_i$ can be either an
insertion, modification, or a deletion. An insertion can occur to the
whole expression or to any sub-expression, and involves replacing $f$
by either $u(f)$ for some unary operation $u$ or by replacing $f$ with
$f \op t$, where $t$ is a constant or a variable instantiation and
$\op$ is some binary operation. A modification uniformly selects a
random operation or operand from the expression and modifies it in
place. Binary (resp.  unary) operations can be modified to be any
other binary (resp. unary) operation. Constants increase or decrease
by one.  Variables either change from an $x_i$ to $y_i$ (or vice
versa) or the index is increased by 1, wrapping back to 1 when
appropriate. Deletion can act on a binary operation or a unary
operation. A unary operation $u(f)$ is replaced by $f$, and a binary
operation $f \op g$ is replaced by a uniform selection of $f$ or $g$.
The distributions on the above choices can be chosen arbitrarily,
however for the present applications the distributions were all taken
to be uniform.

Some of these mutations can transform a completely symmetrized
algebraic expression to one which is not. In this case, first all the
locations for symmetrized operations which would symmetrize a
subexpression are found. Then one is uniformly chosen among these
locations and a uniformly chosen symmetrized operation is applied. If the
resulting expression is still not completely symmetrized the process
is repeated until it is.


Crossover was defined analogous to chromosome sorting in diploid
organisms. Given two BRTSS, one called ``heads'' and the other
``tails'', sample a Bernoulli random variable for each $i$ and choose
the corresponding $\rho_i$ and $\lambda_i$ for the first product of
the crossover. The other product is obtained by using the compliment.
For example, if the sample is $HT$ for $((\lambda_{H,1}, \lambda_{H,2}), (\rho_{H,1}, \rho_{H,2}))$
crossed with $((\lambda_{T,1}, \lambda_{T_2}), (\rho_{T,1},
\rho_{T,2}))$, the resulting BRTSS are 
$((\lambda_{H,1}, \lambda_{T_2}), (\rho_{H,1}, \rho_{T,2}))$ and
$((\lambda_{T,1}, \lambda_{H_2}), (\rho_{T,1}, \rho_{H,2}))$.

In order to avoid overly long BRTSS, it is possible to discount
the fitness of a BRTSS according to its size. Specifically,
rather than
the raw fitness function $F(q)$ one can use $F(q)-\psi S(q)$ where
$S(q)$ is the total size of the BRTSS and $\psi$ is a scaling factor.
It is also possible to have $\psi$ change after a number of
generations of the genetic algorithm.

\subsection{Workflow}
\label{sec:workflow}

Here we describe the strategy for producing high-performance
statistics using the above methodology. First, an objective function
must be chosen which is representative of the problem but which is not
too costly to compute. For instance, to find a statistic which can
differentiate between two distributions on trees, a compromise must be
found for sample size. Too small of a sample may just pick up sampling
differences, yet too large of a sample significantly slows down
computation.

Second, enumeration is used to find a good initial population for the
genetic algorithm. Early efforts demonstrated that the genetic
algorithm was excellent at finding local optima, but that it had
difficulty traversing the whole fitness landscape.
A solution is to start with a diverse population of statistics, which
can be found using the method of Section~\ref{sec:enumeration}. Many
statistics are enumerated and then sorted by their performance; a
selection of the best is then used as an initial population.

Third, the genetic algorithm is run with a variety of parameters and
random seeds. The resulting statistics are then collected and rated
against one another and the best ones found.

The algorithm has been implemented in an \texttt{ocaml} \cite{ocaml}
program; complete source code is available at
http://math.canterbury.ac.nz/matsen/.

\subsection{Overfitting}
\label{sec:overfitting}

The number of verifiably symmetric algebraic expressions--- even of
moderate size and with a small selection of constants--- is enormous.
The number of binary recursive tree shape statistics constructible
with these algebraic expressions is of course significantly larger.
For this reason some caution is needed to avoid ``overfitting'' the
statistical problem at hand. For example, the method described here
can quite easily find a statistic which seems to indicate a
significant difference between two moderately-sized draws from the
same distribution on trees.

This problem can be approached in the following ways. First, in the
applications, we have split the data into ``training'' and
``testing'' data, such that statistics are evolved on the training
data, and then their significance is indicated on the testing data.
If the testing data is of reasonable size, it is unlikely that
observed statistical significance is due to sampling. Second, one can
reduce the overfitting problem by incorporating size into the
fitness function as described above. This tends to keep the
statistics in a more manageable range.  Finally, statistics with
only one recursion are less likely to overfit than those with multiple
recursions; for this reason we have restricted ourselves to the
single-recursion case in the below application.

\section{Application}
\label{sec:applications}

In this section we apply the methods described in the previous chapter
to perform a re-analysis of the results from a recent paper by Blum and
Fran\c{c}ois \cite{blum}. The main purpose of their paper was to
investigate an earlier suggestion by David Aldous that an instance of
his ``beta-splitting'' model might approximate the distribution of
macroevolutionary phylogenetic trees reconstructed from sequence data
\cite{Aldous2001:23}. Blum and Fran\c{c}ois confirm his suggestion,
``that the [imbalance measures] generally agree with a very simple
probabilistic model: Aldous' Branching.'' These models are explained
below. The conclusion of the example application in this paper will be
that although the sampled trees do fit the ``Aldous' Branching'' model
reasonably well in terms of overall balance, it is possible to find a
tree shape statistic which demonstrates a substantial deviation from
the Aldous model.

The ``Aldous' Branching'' model is an instance of a
one-parameter family of models invented by David Aldous called the
``beta-splitting'' models.  These models are simply probability
distributions on trees and are not intended to model any
evolutionary process.  The idea of the beta-splitting
model is to recursively split the taxa into subclades using a
distribution derived from the beta
distribution. More precisely, assuming that a clade has $n$ taxa, the
probability of the split being between subclades of size $i$ and $n-i$
is \[ q_{n,\beta} (i) = C(n;\beta) \, \frac{\Gamma(\beta+i+1)
\Gamma(\beta+n-i+1)} {\Gamma(i+1) \Gamma(n-i+1)} \] where $C(n;\beta)$
is a normalizing constant.  The parameter $\beta$ in Aldous's model
thus determines the overall balance of the trees, such that larger
values of $\beta$ lead to increased balance. The so-called ``equal
rates Markov'' (ERM) model corresponds to $\beta = 0$, and the
``proportional to different arrangements'' (PDA) model results when
$\beta$ is set to $-1.5$. The model when $\beta$ is set to $-1$ is
called the ``Aldous' branching'' model by Blum and Fran\c{c}ois, but we
will simply call it the $\beta=-1$ model.

Blum and Fran\c{c}ois took a sample of trees from the tree database
TreeBASE \cite{treeBASE} and found a maximum-likelihood estimate of
$\beta$ for each of these trees. Because not all of the trees are
binary, they resolved multifurcating nodes (also called polytomies) by
splitting them either via the ERM model (``ERM-solved''
trees) or via the PDA model (``PDA-solved'' trees). They felt that the
inclusion of outgroups might skew the analysis, and thus passed the
trees through an ``automated outgroup removal procedure'' which simply
removes leaves or cherries (subtrees with two leaves) branching off of
the root.

The general strategy taken in this section will be to compare the same
trees used by Blum and Fran\c{c}ois to a sample from the $\beta=-1$
(a.k.a. ``Aldous' branching'')
distribution. Specifically, for each ERM-solved TreeBASE tree in their
set after the outgroup removal procedure, we sample a tree of the same
size from the $\beta = -1$ model.  This provides a paired data
set which is appropriate for paired statistical tests such as the sign
test. As described in Section~\ref{sec:overfitting}, we divide the
data into training and testing subsets. In this case the trees were
numbered starting from zero and the even numbered trees taken for
training and the odd numbered trees taken for testing, resulting in
1032 trees for the training set and 1031 trees for the testing set. 

We first review the statistic used by Blum and Fran\c{c}ois to
compare the TreeBASE trees and the corresponding model
trees. They define
\[
s(T) = \sum_{i \in \mathcal{I}} \log(N_i - 1)
\]
where as before $N_i$ is the number of leaves of the
subtree subtended by internal node $i$. We applied this statistic to
the paired data set, which led to a $p$-value of $0.362$ with the
sign test. Therefore through the eyes of the Blum and Fran\c{c}ois $s$
statistic, the $\beta = -1$ model indeed does a good job of producing
trees similar to those found in TreeBASE.

The goal for the rest of this section will be to find a statistic
which does indicate a significant statistical difference between the
$\beta=-1$ trees and the TreeBASE trees.  Accordingly, the
objective function applied to a chosen statistic was chosen to be the
negative of the logarithm of the $p$-value of the sign test of the
statistic applied to the aligned data.  The recipe from
Section~\ref{sec:workflow} was followed.  In the enumeration phase,
all statistics of length one and size up to five, with constants and
$\lambda_i$ chosen from the set $\{0,1,2\}$, were tested and the best
used as an initial population. The genetic algorithm was run with
population sizes of 50 and 100, 
mutation rate of 20\% per generation, and 1500
generations. 

We will focus on one statistic returned from the algorithm, which will
be called \betast.  The \betast\ statistic has $\lambda = 8$ and
$\rho(x,y) = \left( \log(x+y) \right)^5$. This statistic rejects the
$\beta = -1$ model with a $p$-value of $6.78 \times 10^{-19}$ for the
paired sign test on the testing data.  Therefore this statistic
clearly indicates an important difference between the beta-splitting
and the reconstructed trees.

\begin{figure}
  \begin{center}
    \includegraphics[width=8.8cm]{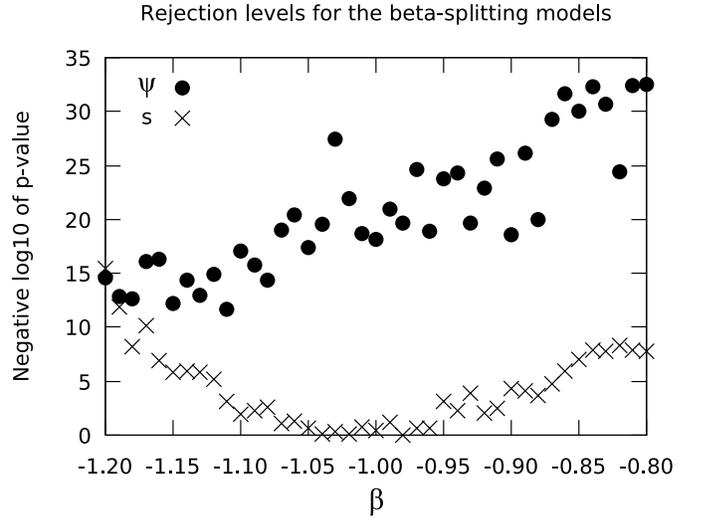}
  \end{center}
  \caption{A comparison of the \betast\ statistic considered in this
  paper and the $s$ statistic of Blum and Fran\c{c}ois.  The $x$ axis
  is the value of $\beta$ used to sample trees, and the $y$ axis is
  the negative base 10 logarithm of the $p$-value of the sign test applied to
  the sample from TreeBASE and samples of the beta-splitting model. Clearly the
  $s$ statistic has low power to distinguish between the two samples
  for a range of $\beta$, while the \betast\ statistic has high power
  within this range.}
  \label{fig:comparison}
\end{figure}

Although the \betast\ statistic was developed in order to
differentiate between the TreeBASE trees and the $\beta=-1$ trees, it
does a good job of differentiating between the sample of TreeBASE
trees and samples from the beta-splitting model for a range of
$\beta$ values.  As seen in Figure~\ref{fig:comparison}, 
the \betast\ statistic rejects the beta-splitting model for a variety
of values of $\beta$ with a very low $p$-value, while the $s$
statistic only rejects the beta-splitting model when $\beta$ is rather
far away from $-1$.

\begin{figure}
  \centering
  \includegraphics[width=8.3cm]{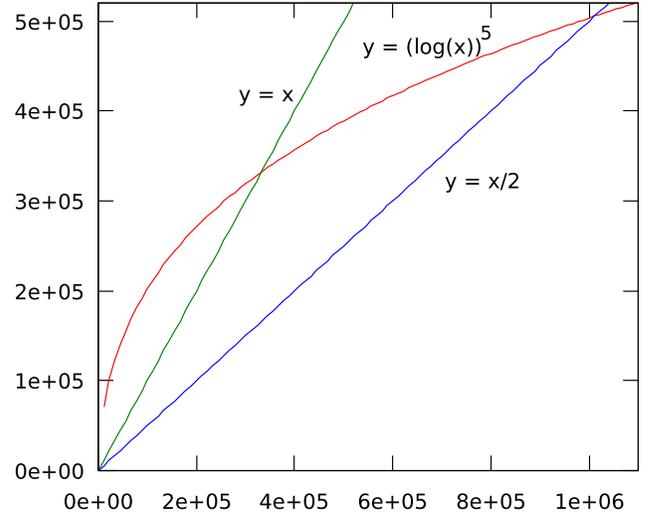}
  \caption{The recursion from \betast.}
  \label{fig:plot}
\end{figure}

We will now sketch some ideas of how the \betast\ statistic might ``work.''
An interesting feature of this statistic is that it converges on
sequences of trees of increasing size satisfying certain conditions.
For example, its value on an infinite balanced tree is approximately
$5.05 \times 10^{5}$, and on an infinite ``comb'' (perfectly
imbalanced) tree its value is
approximately $3.32 \times 10^{5}$. The reasons for this are clear
from Figure~\ref{fig:plot}. The statistic can be evaluated on a
large balanced tree by recursively iterating the function $x \mapsto
\left( \log(2x) \right)^5$; from the plot it is clear that this
recursion will converge to a value slightly more than $5 \times
10^{5}$. For the comb tree the recursion is $x \mapsto \left(
\log(x+8) \right)^5$, and the convergence value can again be estimated
from the plot. 

However, as can be seen from Table~\ref{tab:recur}, the convergence
is not immediate. Furthermore, for small trees the statistic increases
with imbalance (compare the balanced tree of depth three to the comb of
depth seven, each of which have eight leaves), whereas on large trees the statistic increases with
balance. This implies that the exchange of a comb subtree for a balanced
subtree in a small tree can increase the statistic, whereas the same
exchange in a large tree may decrease the statistic. This feature
suggests that the TreeBASE trees may deviate from the ``Markov
branching'' property of the beta-splitting models, which is that the
distribution on subtrees of a given size is independent of the rest of
the tree. 

\begin{table}
  \renewcommand{\arraystretch}{1.3}
  \caption{Calculation of the \betast\ statistic}
  \label{tab:recur}
  \centering
  \begin{tabular}{cccc}
    depth & & balanced & comb \\
    \hline
    0 & & 8 & 8 \\
    1 & & 164 & 94.7 \\
    2 & & 6.52e+03 & 2.04e+03 \\
    3 & & 7.64e+04 & 2.58e+04 \\
    4 & & 2.42e+05 & 1.08e+05 \\
    5 & & 3.85e+05 & 2.09e+05 \\
    6 & & 4.57e+05 & 2.76e+05 \\
    7 & & 4.87e+05 & 3.09e+05 \\
  \end{tabular}
\end{table}

At this point it is important to emphasize that the \betast\ statistic
was invented for the single purpose of distinguishing the
beta-splitting trees from the sorts of trees one finds in TreeBASE.
This statistic was named for convenience only, not to
introduce it into the canon of tree shape statistics. Indeed, one
intent of this paper is to reduce the traditional emphasis on
individual ``general purpose'' tree shape statistics and to focus
instead on creating
statistics for a specific application.  

There will often be many such useful statistics.  For example, note
that a number of statistics appeared on different runs
of the same objective function with similar performance.  Because
the space of algebraic expressions is so large and the fitness
landscape is very ``peaked,'' runs of the genetic algorithm seldom
converge on the same BRTSS when started with a different random seed
or slightly different parameters. 

The results of this section should not be construed as a rejection of
the results or methodology of Blum and Fran\c{c}ois. They found a
value for $\beta$ which does in fact generate the observed level of
overall balance for the TreeBASE trees. However, the above statistic
shows that in this case there is more to tree shape than just overall balance.
The difference between the two perspectives indicates interesting
future directions for research.  For example, is the observed
difference due to reconstruction bias, or is the deviation indicated
by the above statistic an actual feature of macroevolutionary
processes? If the latter,
how can we modify the present models to accommodate the difference?

\section{Conclusions}

In conclusion, we have developed a framework which allows enumeration
of and optimization over a class of tree shape statistics. This class
includes many of the tree shape statistics found in the literature.
A genetic algorithm can be applied in this framework to find
customized tree shape statistics for a certain application. The
methodology is applied in an example case, finding a statistic which
indicates a significant difference between two distributions on trees
which was not previously evident.

Along with this new tool comes a new problem, which is that an
automated system such as the genetic algorithm described above can
create very complex tree shape statistics whose values can be hard to
interpret.  This issue is not problematic from an abstract statistical
viewpoint, however it is comforting to have an intuitive
interpretation of the statistics.  In the sample case above some
intuition was developed about a relatively simple statistic, but it
may not be easy to find an interpretation for a complex one.  It would
be helpful in this regard to be able to derive limiting
distributions for BRTSS applied to a distribution on trees.  It is
possible to do this for certain statistics, such as the number of
cherries \cite{mckenzie-steel} or $I_c$ and $\bar{N}$
\cite{blum-theory}.  The methods used in the latter paper are
applicable to a subclass of the BRTSS, however a substantial amount
of work must be done on a case-by-case basis.

We note that the problem of differentiating two distributions on
combinatorial objects has been approached in a different
fashion by the statistical physics community.  In their case, many
models have been proposed for the growth of social and biological
networks and a goal is to confirm or reject a certain model given some
data.  Analogous to the classical tree shape statistics, individual
means of comparing graphs, such as the diameter or the number of
subgraphs of a specific type, have been described (see, e.g.
\cite{MR1864966}).  A more recent approach is to count in some manner
the number of many different walks on the networks and then feed that
information into a Support Vector Machine \cite{pmid15555081}
\cite{pmid15728374}.  This approach is similar to that described in
the present paper in that machine learning is used to come up with
tests which can distinguish models from data, however the actual
technique is quite different.  Their network approach focuses on local
structure, while the BRTSS in this paper often provide global
information.  Nevertheless, an application of the network approach
might provide some insights in the tree shape setting.

In the future we hope to use the methodology presented in this paper
to expand the applications of tree shape theory in useful directions.
For example, moderately sophisticated models of influenza evolution
are currently being used to elucidate the evolutionary processes which
form the remarkable imbalance of influenza phylogenetic trees
\cite{pmid16723145} \cite{pmid12660783}.  At this point very little of
even the classical tree shape statistics are being applied for
quantitative description.  Another potentially underdeveloped area is
the use of tree shape statistics to detect bias in modern tree
reconstruction methods on real data; a lone article from over 10 years
ago \cite{colless95} forms the complete bibliography in this area.

\section*{Acknowledgments}

FAM was supported by an NSF graduate research fellowship, and would
like to thank and the Allan Wilson Centre for hosting him while much
of this material was developed. A single meeting with Daniel Ford went
a long way towards clarifying and generalizing the set of algebraic
expressions considered here. Martin Willensdorfer introduced the
author to functional programming, which led indirectly to the BRTSS
definition. Michael Blum generously supplied data for the
example application. Steve Evans, Olivier Fran\c{c}ois,
Katherine St. John and Mike Steel provided helpful suggestions on the
research and on earlier versions of this manuscript. Three anonymous
reviewers and the editor provided helpful commentary. The majority of
the computational work was done on the CGR cluster at Harvard.

\bibliography{opt_paper,/home/matsen/cv/my_papers/my_papers.bib,/home/matsen/papers/bibtex_entries}
\bibliographystyle{IEEEtran}

\end{document}